\documentclass[preprint, noinfoline,addressatend,imslayout,a4paper,11pt]{imsart}

\usepackage{url}
\RequirePackage[colorlinks,citecolor=blue,urlcolor=blue]{hyperref}
\usepackage[authoryear]{natbib}

\usepackage{amsthm,amsmath,amsfonts,amssymb}
\usepackage{enumerate}
\usepackage{bm,accents}
\usepackage{graphicx}
\usepackage{color}

\startlocaldefs

\usepackage[plain,noend]{algorithm2e}

\makeatletter
\renewcommand{\algocf@captiontext}[2]{#1\algocf@typo. \AlCapFnt{}#2} 
\def\@algocf@capt@plain{top}
\renewcommand{\algocf@makecaption}[2]{%
  \addtolength{\hsize}{\algomargin}%
  \sbox\@tempboxa{\algocf@captiontext{#1}{#2}}%
  \ifdim\wd\@tempboxa >\hsize
    \hskip .5\algomargin%
    \parbox[t]{\hsize}{\algocf@captiontext{#1}{#2}}
  \else%
    \global\@minipagefalse%
    \hbox to\hsize{\box\@tempboxa}
  \fi%
  \addtolength{\hsize}{-\algomargin}%
}
\makeatother

\def\boxit#1#2#3{\hbox{\vrule width #2               
                     \vtop{%
                             \vbox{\hrule  height #2 \kern#1 
                           \hbox{\kern#1 #3\kern#1}  
                          }%
                     \kern#1 \hrule  height #2   
                  }%
                         \vrule width #2             
                      } %
                  }

\usepackage{latexsym}
\usepackage{epsfig}
\usepackage{subfigure}
\usepackage{lscape}
\usepackage{latexsym}
\usepackage{epsfig}
\usepackage{graphicx}
\usepackage{amsmath,amsfonts,amssymb,bm}

\newtheorem{thm}{Theorem}

\bibpunct{(}{)}{;}{a}{,}{,}
\def\spacingset#1{\renewcommand{\baselinestretch}%
{#1}\small\normalsize} \spacingset{1.6}

\def\bbeta{\beta}
\def\boleta{\mbox{${\eta}$}}

\def\bgamma{\mbox{${\gamma}$}}
\def\sbeta{\mbox{\scriptsize ${\beta}$}}

\def\bxi{\mbox{${\xi}$}}

\def\mpr{\mbox{pr}}

\newcommand{\sumin}{\sum_{i=1}^n}
\newcommand{\sn}{n^{1/2}}
\newcommand{\snm}{n^{-1/2}}
\newcommand{\hpsi}{\hat{\psi}_n}
\newcommand{\tbeta}{\tilde{\bbeta}}
\newcommand{\bo}{{\bbeta^{o}}}
\newcommand{\bZ }{Z}

\newcommand{\bX}{\bar{X}}

\begin{document}
\begin{frontmatter}
%
%

\begin{aug}
\title{A Quantile Regression Model for Failure-Time Data with Time-Dependent Covariates}
\author{\fnms{Malka} \snm{Gorfine}
\ead[label=e1]{gorfinm@ie.technion.ac.il}}
\affiliation{Technion - Israel Institute of Technology }
\printead{e1}

\and
\vspace{-0.2in}

\author{\fnms{Yair} \snm{Goldberg}\ead[label=e2]{ygoldberg@stat.haifa.ac.il} }
\affiliation{University of Haifa}
\printead{e2}

\and

\vspace{-0.2in}
%
\author{\fnms{Yaacov} \snm{Ritov}
\ead[label=e3]{yaacov.ritov@gmail.com}}
\affiliation{The Hebrew University of Jerusalem}
\printead{e3}

\runtitle{Quantile Regression with Time-Dependent Covariates}
\runauthor{Gorfine, Goldberg, Ritov}

\end{aug}









\begin{abstract}
Since survival data occur over time, often important covariates that we wish to consider also change
over time. Such covariates are referred as time-dependent covariates. Quantile regression offers flexible modeling of survival data by allowing the covariates to vary with quantiles.  This paper provides a novel quantile regression model accommodating time-dependent covariates, for analyzing survival data subject to right censoring. Our simple estimation technique assumes the existence of instrumental variables. In addition, we present a doubly-robust estimator in the sense of \citet{robins_recovery_1992}. The asymptotic properties of the estimators are rigorously studied. Finite-sample properties are demonstrated by a simulation study. The utility of the proposed methodology is demonstrated using the Stanford heart transplant dataset.
\end{abstract}

\begin{keyword}
\kwd{quantile regression}
\kwd{time-dependent covariates}
\kwd{survival analysis}
\kwd{instrumental variables}
\end{keyword}
\end{frontmatter}


\section{Introduction}
Quantile regression provides a framework for modeling the relationship between an outcome and covariates using conditional quantile functions~\citep{koenker_regression_1978}. For example, in linear models, quantile regression is a popular alternative to the least-squares approach. Obviously, considering several quantiles of interest provides a more comprehensive statistical analysis than the classical linear regression. With quantile regression methodology, one can estimate the covariates' effect without the assumption that each quantile is related to the covariates in the same fashion as the conditional mean.

For right-censored survival data, quantile regression is emerging as an attractive
alternative to the \citet{cox1972regression} proportional hazards and the accelerated failure time models.
Quantile regression for censored survival data provides a flexible semiparametric modeling tool which does not restrict the variation
of the coefficients for different quantiles, in contrast to the proportional hazards or accelerated failure time models. Hence, quantile regression models are considered robust and flexible in the sense
that they can capture a variety of effects at different quantiles of the
survival distribution. In this work we present a novel model and estimation procedure for right-censored survival data with time-dependent covariates.

Estimation in quantile regression models under right censored survival data with
time-independent covariates has received much attention in the literature
\citep[among others]{powell1984least, powell_censored_1986,ying_survival_1995,mckeague2001median, honore_quantile_2002,portnoy_censored_2003,peng_survival_2008,qian2010censored}. \citet{robins_semiparametric_1992} introduced a class of semiparametric accelerated failure time models for modeling the relationship of survival distribution to time-dependent covariates in the presence of right censoring. They also proposed semiparametric rank estimators for the parameters of the model. Special cases of the Robins-Tsiatis class of models were
already introduced by \citet[Chapter~6]{kalbfleisch_prentice_1980} and \citet[Section~5.2]{cox_analysis_1984}. \citet{Lin1995} derived a semiparametric
inference procedure for the Robins-Tsiatis class of models, along with a
rigorous large-sample theory. This model was also discussed by~\citet{robins_median_1996}, who allowed some dependency between the censoring and some time-dependent auxiliary variables. In a different setting, \citet{bang_median_2002} discussed median regression with censored cost data and time-independent covariates. To the best of our knowledge, none of the published works provide a quantile regression model for censored survival data which handles time-dependent covariates.

As a motivating example, consider the familiar Stanford heart transplant data \citep{crowley1977covariance} where
patients were accepted into the transplant program and then waited until a suitable donor was found.
The survival time is defined as the number of days that elapsed between the date of acceptance and the date
in which each patient was last seen. The main scientific question is whether transplantation prolongs survival.
Let $W$ denote the waiting time from the date of acceptance to the date of heart transplant.
Then, the model includes a time-dependent covariate $X_{1}(t)$ - the transplant status - that takes the value 1 if
$t$ is greater or equal to $W$, and 0 otherwise. Two other covariates of interest are age at transplantation and tissue mismatch score,
which are considered to be prognostic indicators of survival only for patients who received a transplant. Specifically, $X_{2}(t)$ equals the age at transplant for $t \geq W$, and 0 otherwise; and
$X_{3}(t)$ equals the mismatch score at transplant for $t  \geq W$, and 0 otherwise. \citet{Lin1995} analyzed the
data using Cox and accelerated failure time models, each with the above time-dependent covariates.
Their results suggest that transplantation is beneficial for younger patients with lower mismatch score. For example, a patient
transplanted at age of 35 with mismatch score of 0.5, would have lived only 13.7\% of his post-transplantation life
had the patient not received a heart transplantation. In a contrast, the post-transplantation lifetime of a patient aged 53 with mismatch
score of 1.8 would have been \emph{increased} by about 160\% had the operation not been performed.
However, as will be shown in Section~\ref{sec:example}, analysis using the proposed methodology reveals that the effect of age at transplant tends to increase over the quantiles. Hence, the conclusions differs substantially from those obtained using Cox or accelerated failure time models.
For example, among patients of median survival time, the post-transplantation lifetime of a patient aged 53 with mismatch
score of 1.8 would have been \emph {decreased} by about 50\% had the operation not been performed;
for a patient aged 63 it would have been \emph{decreased} only by 5\%; and for a patient aged 73, transplantation is not beneficial.


\section{Model and Estimation}\label{sec:methodology}
\subsection{The model}
Let $\tau$ be a positive random variable and assume for now that it represents the baseline failure time of the investigated phenomenon corresponding to an individual with all covariates equal to zero. Assume $T$ is the actual survival time with survival function $S(\cdot)$,
$\tilde{{X}}(\cdot)$ is a multivariate random process on $[0,\infty)$ of covariates of
dimension $p \times 1$, and $\bgamma^o$ is a $p$-dimensional vector of
coefficients. Following \citet{robins_semiparametric_1992}, it is assumed that if $t$ is the actual
time and $s$ is the baseline time, $ds/dt=\exp\{{\bgamma}^{o'} \tilde{{X}}(t)\}$. Thus,
the observed failure time $T$ is the solution of
\begin{equation}\label{eq:taut0}
\tau=\int_0^T \exp\{{\bgamma}^{o'} \tilde{{X}}(t)\}dt.
\end{equation}
The coefficients in ${\bgamma}^o$ have a direct interpretation in terms of increasing or decreasing the risk. For example, consider the Stanford heart transplant data with
$\tilde{{X}}(t)=I(t \geq W)$, the transplant status, where $I(\cdot)$ denotes the indicator function. A positive coefficient implies that the baseline time is greater than the actual survival time, and thus heart transplant decreases lifespan.
The exponential function in \eqref{eq:taut0} could be replaced by any other positive smooth known function.
In case of time independent covariates, model \eqref{eq:taut0} is reduces to $\tau=T \exp\{{\bgamma}^{o'} \tilde{{X}}\}$.
\citet{robins_semiparametric_1992} and \citet{Lin1995} studied model
\eqref{eq:taut0} for the accelerated failure time model. In the following we provide a new quantile regression model in the spirit of model \eqref{eq:taut0}.

Let ${ X}(t)=\{1,\tilde{ X}(t)'\}'$, $\bar{ X}(t)=\{{ X}(s) , 0 \leq s \leq t \}$ and ${\bar{ X}}=\{{ X}(s) , 0 \leq s < \infty \}$. Assume that the $q$th quantile of the baseline is independent of $X(\cdot)$. That is, there exists a positive real-valued constant $c$ such that the $q$th quantile satisfies
\begin{equation}\label{eq:taulessc}
\mpr \{\tau(q) \leq c | \bar{{X}}\}=q \;\;\;\;\;\;\;\;\; q \in(0,1)\,,
\end{equation}
where
\begin{equation}\label{eq:taut1}
\tau(q)=\int_0^T\exp\{{\bbeta}^{o}(q)' {{X}}(t)\}dt,
\end{equation}
and $\bbeta^o(q)$ is the vector of unknown regression coefficients that represents the effect of the covariates on the $q$th quantile of the survival time. Since the regression coefficient vector includes an intercept term $\beta^o_0(q)$, without loss of generality we may assume $c=1$ and obtain
\begin{equation}\label{eq:tauq}
\mpr \{\tau(q) \leq 1 | \bar{{X}}\}=q \;\;\;\;\;\;\;\;\; q \in(0,1).
\end{equation}

Our quantile regression approach represented by (\ref{eq:tauq}) can be viewed as an extension of the accelerated failure time model of \citet{robins_semiparametric_1992}. In particular, we assumes that the $q$th quantile of
$\tau$ is independent of $\bar{ X}$, and otherwise the distribution of $\tau$ can be dependent of the covariates' process. This model is appropriate when one is interested in the $q$th conditional quantile of $T$. Thus, the proposed model can be considered as a minimal robust alternative to the
model used by \citet[][Eq.~2.2]{robins_semiparametric_1992} in which it is assumed that the distribution of $\tau$, which they refer to as the baseline failure time, is independent of $\bar{ X}$. For simplicity of notation, in what follows, we use $\tau$ and $\bbeta$ instead of
$\tau(q)$ and $\bbeta(q)$, respectively.

\subsection{The estimation procedure}
Define the observed time as $Y = \min(T,C)$, where $C$ is an absolutely continuous right-censoring variable, and let $\Delta = I (T \leq C)$.
In addition, assume the existence of a time-invariant $p$-dimensional
instrumental variable $\tilde{{Z}}\in \mathbb R^p$. ${\tilde{{Z}}}$ can be
${X}(0)$, or any other vector of covariates which is positively
depended on the entire ``treatment" regime
$\bar{{X}}$, but independent of $T$ given $\bar{{X}}$. Let ${Z}=(1,\tilde{{Z}}')'$. In this case, the
observed data consist of $n$ independent and identically distributed replicates of
$\{Y,\Delta,\bar{{X}}(Y),{Z}\}$, denoted by
$\{Y_i,\Delta_i,\bar{{X}}_i(Y_i),{Z}_i\}$, $i = 1,\ldots, n$. The
following estimation procedure uses the assumption that $C$ is independent
of $T$, $\bar{{X}}(T)$ and ${Z}$.

Let $G(\cdot)$ denote the survival function of the censoring variable. Under the
above independent censoring assumption, $E \left\{ \Delta / G(T)|T
\right\}=1$. This motivates us to define our proposed estimator,
$\hat{\bbeta}=(\hat{\beta_0},\hat{\bgamma}')'$ to be $\hat{\bbeta}$ is an approximate solution of
\begin{equation}\label{eq:esteq}
U_n(\bbeta)=n^{-1}\sum_{i=1}^n\frac{\Delta_i {Z}_i}{\hat{G}(Y_i)}
\left(I\left[\int_0^{Y_i} \exp\{ \bbeta' {{X}}_i(t)\} dt  > 1 \right]
-q \right)=0\,,
\end{equation}
where $\hat{G}$ is the Kaplan-Meier estimator of the censoring survival distribution.
Since $U_n$ is discontinuous, (\ref{eq:esteq}) should be replaced in practice by, for example, a minimizer of the Euclidean norm $\|U_n(\bbeta)\|$, yet the solution is not necessarily unique. Various smoothing algorithms can be applied, as often done in quantile regression \citep[among others]{Zang1981, chen2005computational}. In the simulation setting and the Stanford heart transplant data analysis, we approximated $I(x>0)$ by $1/\{1+\exp(-ax)\}$
with large value of $a$, and used the Euclidean norm.

For the variance estimator of $\hat{\bbeta}$,
the weighted bootstrap approach is adopted. Specifically, at each bootstrap iteration $b$, $b=1,\ldots,B$, generate $n$ random positive weights, $\omega^{(b)}_1,\ldots,\omega^{(b)}_n$, from the unit exponential distribution; compute the weighted Kaplan-Meier estimator of the censoring survival distribution denoted by $\hat{G}^{(b)}$; and compute the weighted estimator, $\hat\bbeta^{(b)}$, by replacing in $\|U_n(\bbeta)\|$ the function $\hat{G}$ by the function $\hat{G}^{(b)}$ and each $\Delta_i$ by $\omega^{(b)}_i\Delta_i$, $i=1,\ldots,n$. The sample variance of $\hat\bbeta^{(1)},\ldots,\hat\bbeta^{(B)}$ provides a reasonable estimate of the variance of $\hat{\bbeta}$, as shown in Section 5. This procedure can be theoretically justified by Corollary 13.8 of \citet{Kosorok08} which discusses weighted bootstrap and estimating equations involving sums of dependent random variables.

\section{Asymptotic Properties}\label{sec:asymptotics}
Denote the martingale of the censoring time by $M_{Gi}(t)=N_{Gi}(t)-R_{i}(t)
 \Lambda_G(t)$ with respect to the history $\mathcal{F}_{Gi}(t)=\{
N_{Gi}(u), I(Y_i \geq u); 0 \leq u \leq t\}$ where $i=1,\ldots,n$,
$N_{Gi}(t)=I(Y_i \leq t)(1-\Delta_i)$, $R_{i}(t)=I(Y_i \geq t)$, and $\Lambda_G(t)=-\log G(t)$. Let
$$
A(\bbeta)=-E\left[
{Z}
\frac{\partial \theta(\bar{{X}},\bbeta)}{\partial
\bbeta} f\{\theta(\bar{{X}},\bbeta)|\bar X\}\right]\,,
$$
where $\theta(\bar{{X}}_i,\bbeta)$ is defined as the value
of $\theta$ such that $1=\int_0^\theta \exp\{\bbeta' {X}_i(s)\} ds$,
$f(\cdot|\cdot)$ is the conditional density of $T$ given $\bar{{X}}$ which we assume it exists,
and $\partial \theta(\bar{{X}}_i,\bbeta)/\partial \bbeta$
is a $1 \times (p+1)$ vector such that its $j$th component equals
$$
\left[ \int_0^{\theta(\scriptsize \bar{{X}}_i,\sbeta)}
\exp\{\bbeta' {X}_i(t)\} X_{ij}(t) dt \right]^{-1} \;\;\;\;
j=1,\ldots,p+1.
$$
Finally, let
$$
u(\bbeta^o,v)=\lim_{n \rightarrow \infty} n^{-1}\sum_{i=1}^n\frac{\Delta_i
{Z}_i}{G(Y_i)}
\left(I\left[\int_0^{Y_i} \exp\{\bbeta^{o'} {X}_i(t)\} dt  > 1
\right] -q \right)R_i(v)\,.
$$
Denote $R(v)=\sum_{i=1}^n R_i(v)$ and $r(v)=\lim_{n \rightarrow \infty}
n^{-1}R(v)$.

We need
the following regularity conditions for the asymptotic results presented in Theorem \ref{thm1} below:

\begin{enumerate}
\renewcommand{\labelenumi}{(A\arabic{enumi})}
\item \label{as:A1}
 $\bar{{X}}$ and ${Z}$ are uniformly bounded.
\item $\bbeta^o$ lies in the interior of a bounded convex region
$\mathcal{B}$.
\item There exists a constant $\tilde{y}>0$ such that $\mpr (Y>\tilde{y})>0$.

\item \label{as:A5} $\det \left\{ A(\bbeta^o) \right\} \neq 0$.
\end{enumerate}
Assumption~\ref{as:A5} holds if $\bar{{X}}$ and ${Z}$ are positively dependent.

\begin{thm}\label{thm1}
Suppose the model given by (\ref{eq:taut1}) and (\ref{eq:tauq}) holds, and let $\hat{\bbeta}$ be a minimizer of $\|U_n(\bbeta)\|$. If Assumptions \ref{as:A1}--\ref{as:A5} hold, then as $n \rightarrow \infty$:
\begin{enumerate}
\item  $\hat{\bbeta}$ converges to $\bbeta^o$ almost surely.
\item $n^{1/2}U_n(\bbeta^o)$ is asymptotically mean zero multivariate normal vector with
covariance matrix $\Psi=E(\boleta_1 \boleta_1')$,
where for $i=1,\ldots,n$
$$
\boleta_{i}=\frac{\Delta_i {Z}_i}{G(Y_i)}
\left(I\left[\int_0^{Y_i} \exp\{\bbeta^{o'} {X}_i(t)\} dt  > 1
\right] -q\right)  - \int_0^\infty \frac{u(\bbeta^o,v)}{r(v)}dM_{Gi}(v).
$$
\item $n^{1/2}(\hat{\bbeta}-\bbeta^o)$ is asymptotically mean zero multivariate normal vector with covariance matrix
$A(\bbeta^o)^{-1} \Psi A(\bbeta^o)^{-1}$.
\end{enumerate}
\end{thm}
The proof is given in the Appendix.

\section{Augmentation-Based Estimator}\label{sec:augmentation}
In Section~\ref{sec:methodology} we discussed the estimator $\hat\bbeta$, which is obtained as an approximate zero to the estimating equation~\eqref{eq:esteq}. We note that this estimating equation is constructed as a sum of expressions that are different from zero only for indices of observations that are not censored. Thus, the only information obtained from the censored observations is in estimating the $G$, the survival function of the censoring variable. Following the methodology of \citet{robins_recovery_1992}, we propose an augmentation-based estimator that takes into account the censored observations. The estimator that we present is an approximate zero of an estimating equation which is obtained from~\eqref{eq:esteq} by adding an additional augmentation expression. As will be explained in detail below, this augmentation expression is obtained by first positing a model for the distribution of $\{\bZ,T,\bX(T)\}$ and then calculating expectations with respect to this model. We refer to the obtained estimator $\tilde\bbeta$ as the augmentation-based estimator.

As discussed in \citet{robins_recovery_1992} and in more detail in~\citet{laan_unified_2003} and \citet{Tsiatis2006Semiparametric}, when the augmentation expression is chosen well, the advantages of the augmentation-based estimator are two-fold. First, the estimator is consistent when either the censoring distribution does not depend on the covariates, or the posited model for $\{\bZ,T,\bX(T)\}$ is correct. For this reason, this estimator is referred to as a doubly-robust estimator. Second, when the censoring distribution does not depend on the covariates and the posited model for $\{\bZ,T,\bX(T)\}$ is correct, the augmented-based estimator $\tilde \bbeta$ has a smaller asymptotic variance than $\hat\bbeta$. One disadvantage of the proposed augmentation-based estimator is that one needs to posit a model for the distribution of $\{\bZ,T,\bX(T)\}$, and calculate expectations according to this model. This can be computationally demanding. Another potential disadvantage is that when the posited model is chosen poorly, the asymptotic variance of the estimator can actually grow. In the following, we will present the proposed estimator and discuss its asymptotic properties.

Let $H_i=\{Y_i,\Delta_i,\bX_i(Y_i),\bZ_i\}$ and let $H_i(r)=\{\bZ_i,\bX_i(r)\}$ if $r<Y_i$, and $H_i$ otherwise.
Let $\mathcal{P}=\{p(h;\psi);\psi\in\mathbb{R}^q\}$ be a posited finite-dimensional statistical model of the distribution of $\{\bZ,T,\bX(T)\}$. Let $\hpsi$ be the maximum likelihood estimator for this model and let $\psi^*$ be its limit. In the following we assume that $\sn(\hpsi-\psi^*)=O_p(1)$. For modeling of distributions and estimation in this setting we refer the reader to \citet[Chapter~3.5]{laan_unified_2003}.


Define
\begin{align*}
  Q\{s,\bbeta,\psi,H(s)\}=E\left\{\left. {\bZ}\left( I\left[\int_0^{T} \exp\{ \bbeta' {{X}}(t)\} dt  > 1 \right]-q\right) \right|T\geq s, H(s),\psi\right\}
\end{align*}
and $d\hat{M}_{Gi}(t) = dN_{Gi}(t)-R_i(t)d\hat{\Lambda}_{Gi}(t)$.
Let
\begin{eqnarray}\label{eq:esteqdr}
&&U_n^{DR}(\bbeta)=\\
&&\quad n^{-1}\sum_{i=1}^n\left\{\frac{\Delta_i \bZ_i}{\hat{G}(Y_i)}
\left(I\left[\int_0^{Y_i} \exp\{ \bbeta' {{X}}_i(t)\} dt  > 1 \right]
-q \right)+\int_0^{\infty}Q\{t,\bbeta,\hpsi,H_i(t)\} \frac{d\hat{M}_{Gi}(t)}{\hat{G}(t)} \right\}\nonumber
\end{eqnarray}
and let $\tbeta$ be an approximate zero of $U_n^{DR}$. We replace an earlier assumption that the censoring is independent of the failure time and covariates, i.e, missing completely at random (Section 2.2), with the relaxed assumption of missing at random.

\begin{enumerate}
\renewcommand{\labelenumi}{(A\arabic{enumi})}
\addtocounter{enumi}{4}
\item \label{as:A6}
The data is coarsened at random. In other words, the hazard of the censoring variable $C$ at time $v$, given the full data $\{\bZ,T,\bX(T)\}$ and $T\geq v$, is a function only of the observed data $\{\bZ,T\geq v,\bX(v)\}$.
\end{enumerate}
\begin{thm}\label{thm2}
Let the model given by (\ref{eq:taut1}) and (\ref{eq:tauq}) hold, and let $\tilde{\bbeta}$ be a minimizer of $\|U_n^{DR}(\bbeta)\|$. Then, under Assumptions \ref{as:A1}--\ref{as:A6}, as $n \rightarrow \infty$,
\begin{enumerate}
\item $\tbeta$ converges to $\bbeta^o$ almost surely if either the posited model for the distribution of $\{\bZ,T,\bX(T)\}$ holds or the censoring variable is independent of the failure time and covariates.
\end{enumerate}
Moreover, if the censoring variable is independent of the failure time and covariates, then
\begin{enumerate}
\addtocounter{enumi}{1}
\item $n^{1/2}U_n(\bbeta^o)$ is asymptotically mean zero multivariate normal vector with
covariance matrix $\Psi^{DR}=E(\bxi_1 \bxi_1')$,
where for $i=1,\ldots,n$,
$$
\bxi_{i}=\int_0^\infty\left[Q\{t,\bo,\psi^*,H_i(v)\}  -\frac{E\{m(H,\bo)I(T\geq v)\}}{S(v)}\right]dM_{Gi}(v) + \Delta_i \frac{m(H_i,\bbeta)}{G(Y_i)}
$$
and
\begin{align}\label{eq:m}
  m(H_i,\bbeta)\equiv {Z}_i\left(I\left[\int_0^{T_i} \exp\{\bbeta' {X}_i(t)\} dt  > 1 \right]
-q \right)\,.
\end{align}

\item $n^{1/2}(\tbeta-\bbeta^o)$ is asymptotically mean zero multivariate normal vector with covariance matrix
$A(\bbeta^o)^{-1} \Psi^{DR} A(\bbeta^o)^{-1}$.
\end{enumerate}
\end{thm}
The proof is given in the Appendix.

\section{Simulation Study}\label{sec:simulation}
We consider a situation in which there are stepwise time process covariates. Specifically, for $i=1,\ldots,n$, $X_{i0}(t) \equiv 1$, $X_{i1}(t)= S_{i1}  I\{t\in(W_{i1},W_{i2}]\}$ and $X_{i2}(t) = S_{i2} I(t>W_{i2})$, so that $X_{i1}(t)$ and $X_{i2}(t)$
represent levels of, for example, a drug given to patient $i$ at time $t \in [0,\infty)$. The instrumental variable is a bivariate vector $(Z_{i1},Z_{i2})'$, such that $Z_{i1}$ and $Z_{i2}$ are independent unit exponential random variables, $i=1,\ldots,n$. Two scenarios are considered for the drug dosage change points of each subject $i$, $W_{i1}$ and $W_{i2}$: (i) fixed changepoints $(W_{i1},W_{i2})=(0.6,0.9)$; and (ii) $W_{i1}$ and $W_{i2}-W_{i1}$ are independent and exponentially distributed with mean 0.25. 
The actual drug dosages of subject $i$ at the intervals $(W_{i1},W_{i2}]$ and $(W_{i2},\infty]$, are determined by $S_{ij}=V_{ij}+Z_{ij}/2$, $j=1,2$, such that $V_{ij}$ are independent gamma random variables with shape 4 and scale 0.2.
The intrinsic time of subject $i$ is defined as $\tau_i = \tilde{\tau}_i \exp(\beta_0)$ such that $\tilde{\tau}_i$ is gamma distributed with shape $S_{i1}$ and scale $1/c(S_{i1})$, where $c(x)$ is the median of the gamma distribution with shape $x$ $(x>0)$ and scale 1. Hence it is easy to verify that $\mpr (\tau_i \leq 1 | S_{i1}) = 0.5$, $i=1,\ldots,n$, and~\eqref{eq:taut1} holds with $q=0.5$. Finally, by solving~\eqref{eq:taulessc}, the actual failure time of subject $i$ is given by
\begin{eqnarray}
T_{i} =  \left\{
\begin{array}
{l@{\quad \mbox{if} \quad}l}
\tau_i & \tau_i \leq W_{i1} \\
W_{i1}+(\tau_i-W_{i1}){\exp(-\beta_1 S_{i1})} & W_{i1}<\tau_i<W_{i1}+A_i\\
W_{i2}+(\tau_i-W_{i1}-A_i)\exp(-\beta_2 S_{i2}) &
\tau_i\geq W_{i1}+A_i
\end{array}
\right.,
\end{eqnarray}
where $A_i=(W_{i2}-W_{i1})\exp(\beta_1 S_{i1})$.
The censoring times, $C_i$, $i=1,\ldots,n$, are assumed to be exponentially distributed with rate defined by the desired censoring rate.

For the solution of Eq.~(\ref{eq:esteq}), we used $1/\{1+\exp(-\alpha x)\}$, with $\alpha=20$, as a smooth approximation to the step function $H(x)=I(x>0)$. Obviously, this approximation is differentiable at any point and to any order, so that well-known algorithms, such as Broyden \citep{dennis1996numerical}, can be easily used.

Tables \ref{table_a} and \ref{table_b} summarize the results for $n=200$, $500$, and $1000$, and two censoring rates. The true regression coefficient vector equals $\bbeta^o=(-1,1,1)'$. For each case we present the empirical mean, median, standard deviation (SD), interquartile distance (IQ-SD), and the coverage rate of a 95\% weighted bootstrap confidence interval using 500 bootstrap samples. The IQ-SD is defined as the interquartile range divided by 1.349, where 1.349 is the ratio between the interquartile range and the standard deviation for a normal distribution. Clearly, the median and the interquartile distance are robust measures to outliers for the location and dispersion, respectively, and therefore might provide additional important information beyond the mean and SD.
Results of fixed and random change-points are presented in Tables \ref{table_a} and \ref{table_b}, respectively.
The results are based on 1000 Monte Carlo trials. It is evident that the proposed estimation procedure performs very well in terms of bias and that the empirical coverage rates are reasonably close to the nominal level.

\begin{table}
\caption{Simulation results: $\bbeta^o=(-1,1,1)'$; $W=(0.6,0.9)'$ }{
\begin{tabular}{crccccccccccc}
& &\multicolumn{5}{c}{20\% censoring rate  }&& \multicolumn{5}{c}{40\% censoring rate  }\\
parameter & $n$     & mean   & median & SD    & IQ-SD & 95\%CI & & mean   & median & SD    & IQ-SD & 95\%CI \\
$\beta_0^o$ & 200  & -1.089 & -0.995 & 0.586 & 0.510  & 0.972  &  & -1.038 & -0.898 & 0.684 & 0.586 & 0.939  \\
& 500  & -1.035 & -0.988 & 0.348 & 0.309 & 0.972  &  & -0.998 & -0.94  & 0.423 & 0.359 & 0.945  \\
& 1000 & -1.009 & -0.991 & 0.219 & 0.217 & 0.952  &  & -0.976 & -0.966 & 0.267 & 0.245 & 0.939  \\
$\beta_1^o$ & 200  & 1.012  & 0.998  & 0.343 & 0.287 & 0.970   &  & 0.965  & 0.958  & 0.413 & 0.333 & 0.936  \\
& 500  & 1.011  & 0.995  & 0.185 & 0.175 & 0.970   &  & 0.984  & 0.968  & 0.228 & 0.212 & 0.947  \\
& 1000 & 0.999  & 0.995  & 0.121 & 0.120  & 0.961  &  & 0.978  & 0.971  & 0.150  & 0.141 & 0.936  \\
$\beta_2^o$ & 200  & 1.078  & 0.967  & 0.831 & 0.466 & 0.957  &  & 1.001  & 0.872  & 1.068 & 0.572 & 0.931  \\
 & 500  & 1.036  & 0.984  & 0.497 & 0.290  & 0.957  &  & 1.002  & 0.930   & 0.619 & 0.333 & 0.940   \\
& 1000 & 1.007  & 0.979  & 0.312 & 0.196 & 0.948  &  & 0.979  & 0.953  & 0.386 & 0.226 & 0.931  \\
\end{tabular}}\label{table_a}
\end{table}

\begin{table}
\caption{Simulation results: $\bbeta^o=(-1,1,1)'$; $W_1,W_2-W_1\sim \mbox{Exp}(4)$ }{
\begin{tabular}{crccccccccccc}
& &\multicolumn{5}{c}{18\% censoring rate  }&& \multicolumn{5}{c}{34\% censoring rate  }\\
parameter & $n$    & mean   & median & SD    & IQ-SD & 95\%CI &  & mean   & median & SD    & IQ-SD & 95\%CI \\
$\beta_0^o$ & 200  & -1.089 & -0.995 & 0.586 & 0.510  & 0.972  &  & -1.038 & -0.898 & 0.684 & 0.586 & 0.939  \\
& 500  & -1.035 & -0.988 & 0.348 & 0.309 & 0.972  &  & -0.998 & -0.940  & 0.423 & 0.359 & 0.945  \\
& 1000 & -1.009 & -0.991 & 0.219 & 0.217 & 0.952  &  & -0.976 & -0.966 & 0.267 & 0.245 & 0.939  \\
$\beta_1^o$ & 200  & 1.012  & 0.998  & 0.343 & 0.287 & 0.970   &  & 0.965  & 0.958  & 0.413 & 0.333 & 0.936  \\
& 500  & 1.011  & 0.995  & 0.185 & 0.175 & 0.970   &  & 0.984  & 0.968  & 0.228 & 0.212 & 0.947  \\
& 1000 & 0.999  & 0.995  & 0.121 & 0.120  & 0.961  &  & 0.978  & 0.971  & 0.150  & 0.141 & 0.936  \\
$\beta_2^o$ & 200  & 1.078  & 0.967  & 0.831 & 0.466 & 0.957  &  & 1.001  & 0.872  & 1.068 & 0.572 & 0.931  \\
& 500  & 1.036  & 0.984  & 0.497 & 0.290  & 0.957  &  & 1.002  & 0.930   & 0.619 & 0.333 & 0.940   \\
& 1000 & 1.007  & 0.979  & 0.312 & 0.196 & 0.948  &  & 0.979  & 0.953  & 0.386 & 0.226 & 0.931  \\
\end{tabular}}
\label{table_b}
\end{table}

\section{Example - Stanford Heart Transplant Data}\label{sec:example}
\begin{figure}[!t]
\begin{center}
\includegraphics[width=7.0cm, height=7.0cm]{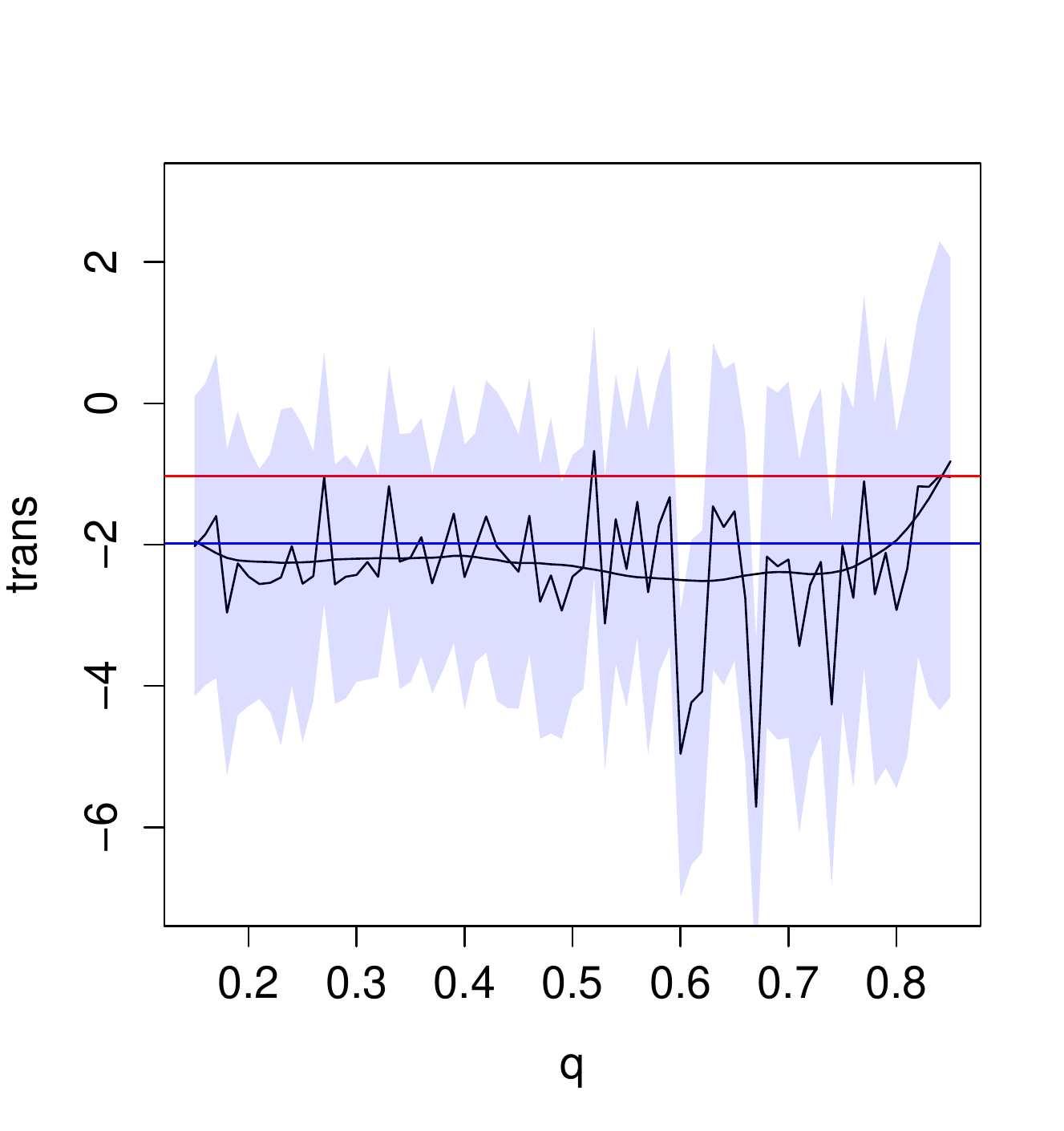}
\includegraphics[width=7.0cm, height=7.0cm]{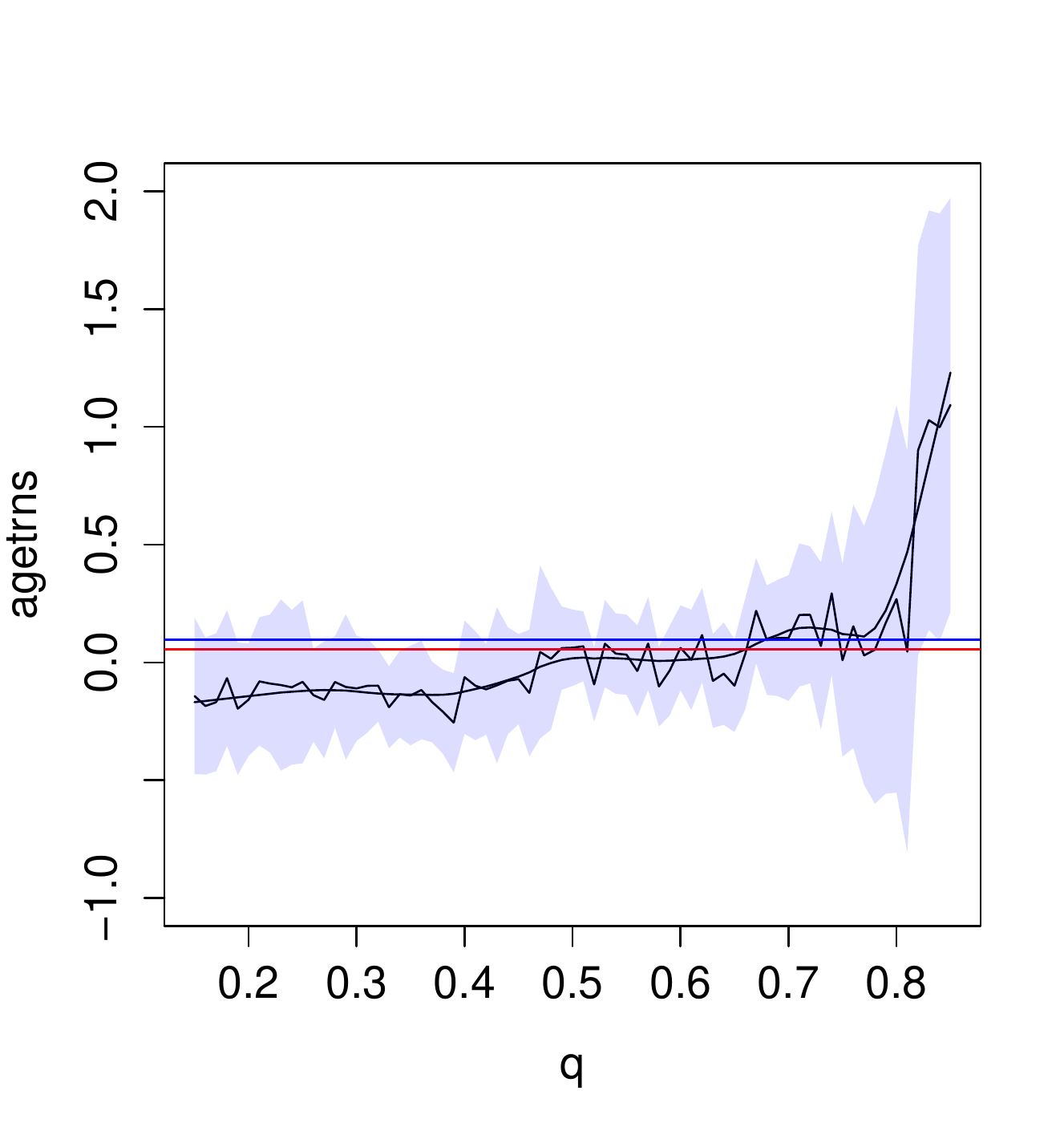}
\includegraphics[width=7.0cm, height=7.0cm]{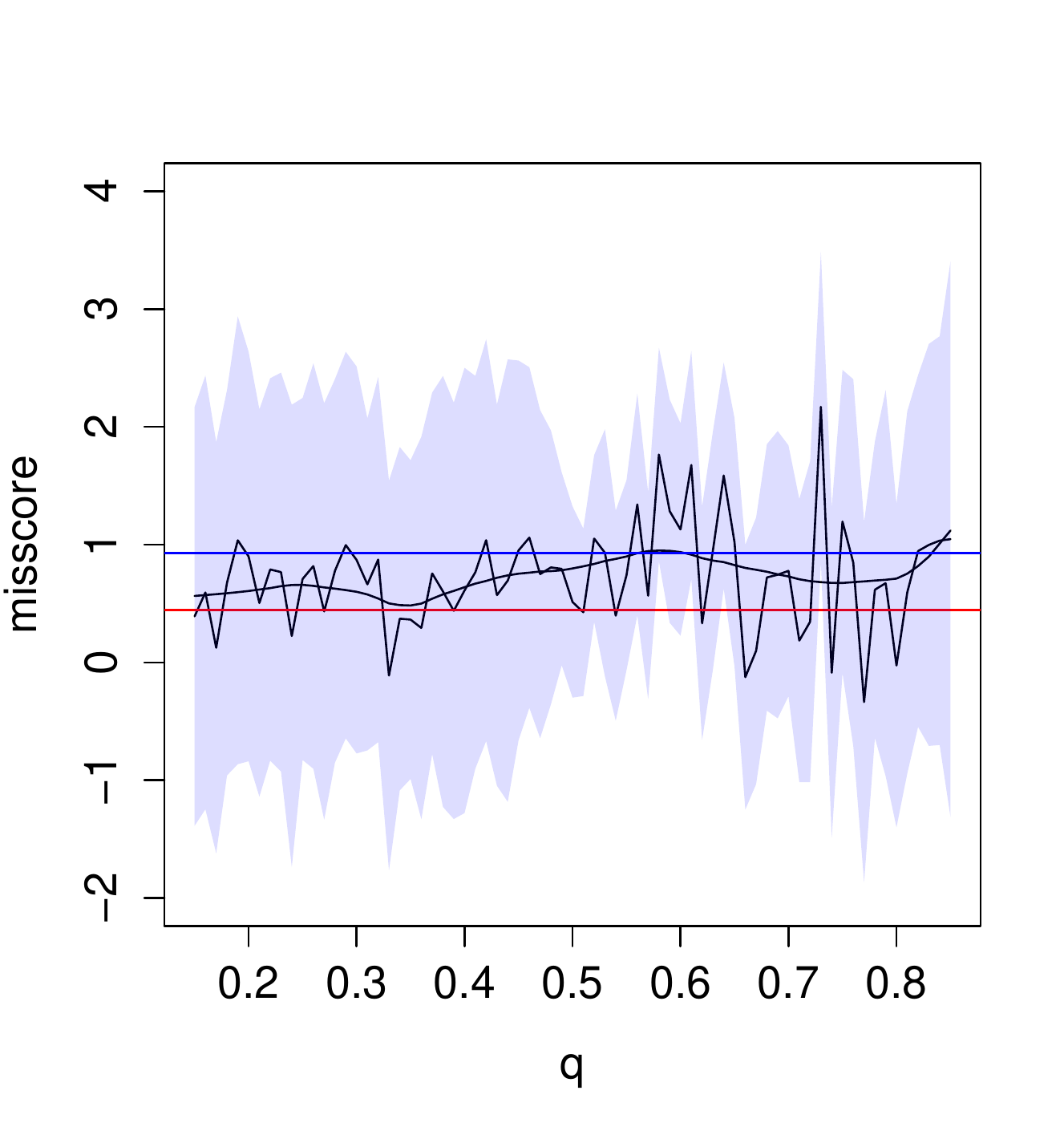}
\caption{Analysis of Stanford heart transplant data: red line - Cox model, blue line - accelerated failure time model, black line and curve - the proposed estimator and its smoothed curve. The shaded area - 95\% bootstrap point-wise confidence interval based on the proposed quantile-based methodology and 500 bootstrap samples.\label{fig1}}
\end{center}
\end{figure}

We illustrate our model and estimation procedure with the familiar Stanford heart transplant data \citep{crowley1977covariance}, available in the package \texttt{survival} of~R. We contrast our model with the Cox proportional hazards model and accelerated failure time model, both with time-dependent covariates. Patients were accepted into the transplant program, and then waited until a suitable donor was found. The survival time is defined as the number of days elapsed between the date of acceptance and the date on which the patient was last seen. The goal of the present is to explore the simultaneous effect of several covariates on survival. In particular, we check whether transplantation prolongs survival. The following analysis is based on 99 patients, 28 of whom were censored as of the closing date. Following \citet{Lin1995}, we consider three time-dependent covariates: $X_{i1}$ - transplant status, $X_{i2}$ - age at transplant, and $X_{i3}$ - mismatch score, $i=1,\ldots,n$. Specifically, let $W_i$ denote the waiting time from the date of acceptance to the date of transplant, of patient $i$. Then, $X_{i1}(t)=I(t \geq W_i)$,
\begin{eqnarray}\nonumber
X_{i2}(t) =  \left\{
\begin{array}
{l@{\quad \mbox{if} \quad}l}
0 & t < W_{i} \\
\mbox{age at transplant minus 35}  & t \geq W_i\\
\end{array}
\right.
\end{eqnarray}
and
\begin{eqnarray}\nonumber
X_{i3}(t) =  \left\{
\begin{array}
{l@{\quad \mbox{if} \quad}l}
0 & t < W_{i} \\
\mbox{mismatch score minus 0.5}   & t \geq W_i \\
\end{array}
\,.\right.
\end{eqnarray}
For applying the proposed quantile model and estimation procedure, we let $\tilde{Z}_{ij}=X_{ij}(Y_i)$, $j=1,2,3$, $i=1,\ldots,n$,
and minimize the Euclidean norm $\| U_n(\bbeta) \|$ after
replacing the indicator function $I(x>0)$ by $\{1+\exp(-100x)\}^{-1}$. Table~\ref{table2} reports the results based on Cox and accelerated failure time models, each with time-dependent covariates, as reported in \citet{Lin1995}. Table~\ref{table3} reports the point estimates and the weighted bootstrap 95\% confidence intervals based on our regression model, for the quartiles. It is evident
that under the Cox regression analysis, transplantation status and age at transplantation are significant, but mismatch score is not. The analysis of the accelerated failure time model provides stronger effects of transplant status and age at transplant, compared to the Cox regression analysis. Our results reveal even stronger effect of transplant status, that varies across patients at different survival stages. The mismatch score effect, although not significant, also indicate for changes as a function of survival stage. Figure~\ref{fig1} also demonstrates that the three models, Cox proportional hazards, accelerated failure time, and quantile regression, are telling us different stories. The detailed discussion of these results, provided in the Introduction, reveals that the three models can lead to dramatically different conclusions. Due to the flexibility of the quantile regression model over the Cox proportional hazards and the accelerated failure time models, along with the meaningful results of the quantile regression analysis, we conclude that the quantile analysis of this dataset is the more reliable.

\begin{table}
\caption{Lin and Ying's Regression analyses of the Stanford heart transplant data}{
\begin{tabular}{lcccc}
& \multicolumn{2}{c}{Cox model} & \multicolumn{2}{c}{Accelerated life model} \\
      Covariate     & estimate  & Wald statistic* & estimate & test statistic* \\
Transplant status &          -1.031  & 4.56 & -1.986 & 4.85 \\
Age at transplant minus 35 &         0.055 & 5.94 & 0.096 & 8.88 \\
Mismatch score minus 0.5  &         0.445 & 2.52 & 0.930 & 2.02 \\

\end{tabular}}
\label{table2}
\\{*Compared against chi-squared distribution with 1 degree of freedom.}

\end{table}

\begin{table}
\def~{\hphantom{0}}
\caption{Quantile Regression analysis of the Stanford heart transplant data}{
\begin{tabular}{clcc}
$q$& Covariate & point estimate  & 95\% CI \\
1/4 & intercept         &              -4.178  &  (-4.558 , -3.798) \\
& Transplant status &              -2.553   & (-4.220 , -0.885) \\
& Age at transplant minus 35 &        -0.083   & (-0.475 , 0.310) \\
& Mismatch score minus 0.5  &         0.708    & (-1.237 , 2.654) \\
1/2 & intercept         &              -4.373   & (-5.348 , -3.397) \\
& Transplant status &              -2.452   & (-4.267 , -0.637) \\
& Age at transplant minus 35 &        0.062    & (-0.114 , 0.238) \\
& Mismatch score minus 0.5  &         0.513  & (-0.346 , 1.371) \\
3/4 & intercept         &              -3.623    & (-5.310,-1.936) \\
& Transplant status &              -2.013     & (-4.425,0.399) \\
& Age at transplant minus 35 &        0.009    & (-0.437,0.455) \\
& Mismatch score minus 0.5  &         1.196     & (-0.739,2.466) \\
\end{tabular}}
\label{table3}
\end{table}

\section{Summary}\label{sec:discussion}
We presented a novel model for quantile regression with time-dependent covariates where the data is subject to right censoring. The estimation procedure of Section~\ref{sec:methodology}, which assumes independent censoring, can be easily applied and possesses good asymptotic  properties. Our numerical studies show that the empirical bias is very small and the coverage rates are fairly close to the nominal level, even with moderate sample size and substantial censoring rates.
We showed that this estimator can be improved by the consistent and asymptotically normal doubly-robust estimator of Section~\ref{sec:augmentation}. While we find this augmented-based estimator theoretically interesting, in practice, it might be difficult to estimate the function $Q$ which requires modeling the distribution of  $\{T,H(s); s \leq T \}$.

The Cox regression model is considered as a cornerstone of modern survival analysis. One of its strengths is the ability to encompass covariates that change over time, due to the theoretical foundation of martigales. On the other hand, a strong and quite apparent violation of the proportional hazards assumption occurs if for two different covariate vectors, their survival functions, or equivalently the conditional quantiles, do cross. The accelerated failure time class of models with time-dependent covariates \citep{robins_semiparametric_1992} is a useful alternative to the Cox regression model,
and the quantile regression model proposed in this work can be viewed as a flexible extension of the Robins-Tsiatis accelerated failure time class of models.

\section*{Acknowledgement}
The work of M.\ Gorfine is supported in part by NIH grant P01CA53996. Y.\ Goldberg was funded in part by ISF grant 1308/12. Y.\ Ritov was supported in part by an ISF grant.

\appendix
\section*{Appendix 1}
\subsection*{Proof of Theorem~\ref{thm1}}
\noindent\emph{ Part 1:}
Let
\begin{align}\label{eq:Un_tilde}
\tilde{U}_n(\bbeta)=\frac{1}{n}\sum_{i=1}^n\frac{\Delta_i
{Z}_i}{G(Y_i)}\left(
\mpr \left[\int_0^{T_i} \exp\{\bbeta' {X}_i(t)\} dt > 1  \right] -q
\right),
  \end{align}
and note that for some $\delta>0$, $\sup_{0 \leq y \leq \tilde{y}} |\hat{G}(y)-G(y)|=o(n^{-1/2+\delta})$ almost surely as $n \rightarrow \infty$ \citep[page~418]{Cosorgo_1983}.
We look at $U_n(\bbeta)-\tilde{U}_n(\bbeta)$ by adding and subtracting
$$\frac{\Delta_i {Z}_i}{G(Y_i)}
\left(I\left[\int_0^{Y_i} \exp\{\bbeta' {X}_i(t)\} dt  > 1 \right]
-\mpr \left[\int_0^{T_i} \exp\{\bbeta' {X}_i(t)\} dt > 1  \right]
\right).$$
Thus, we get
\begin{eqnarray}
U_n(\bbeta)-\tilde{U}_n(\bbeta)&=&\frac{1}{n}\sum_{i=1}^n\frac{\Delta_i
{Z}_i}{G(Y_i)}\left(
I\left[\int_0^{T_i} \exp\{\bbeta' {X}_i(t)\} dt > 1  \right]
\right.  \nonumber \\
&& \left. -\mpr \left[\int_0^{T_i} \exp\{\bbeta' {X}_i(t)\} dt > 1
\right]
\right)+o(n^{-1/2+\delta}) \nonumber
\end{eqnarray}
almost surely.
Now, define the following class of functions where $G(\cdot)$ is bounded
away from 0,
$$
\mathcal{F}=\left\{\frac{\Delta_i {Z}_i}{G(Y_i)}\left(
I\left[\int_0^{T_i} \exp\{\bbeta' {X}_i(t)\} dt > 1  \right] -\mpr
\left[\int_0^{T_i} \exp\{\bbeta' {X}_i(t)\} dt > 1  \right]\right)
\;\;, \;\; \bbeta \in \mathcal{B}  \right\}.
$$
The stochastic process
$$\frac{\Delta_i {Z}_i} { G(T_i)} \left(
I\left[\int_0^{T_i} \exp\{\bbeta' {X}_i(t)\} dt > 1  \right] -\mpr
\left[\int_0^{T_i} \exp\{\bbeta' {X}_i(t)\} dt > 1
\right]\right)$$
is cadlag nondecreasing, the class of indicator functions is Donsker, and
$\bar{{X}}$, ${Z}$ and $G(\cdot)$ are uniformly bounded. It follows, therefore, from example 2.11.16 of \citet[page 215]{VW96} that the class $\mathcal{F}$ is Donsker, and thus
Glivenko-Cantelli applies. Hence,
$\sup_{\bbeta \in \mathcal{B}}||
U_n(\bbeta)-\tilde{U}_n(\bbeta)||=o(n^{-1/2+\delta})$ almost surely as $n
\rightarrow \infty$. Also, note that $\tilde{U}_n(\bbeta^o)=0$ and
$$
\frac{\partial \tilde{U}_n(\bbeta)}{\partial \bbeta} = -\frac{1}{n}\sum_{i=1}^n
\frac{\Delta_i
{Z}_i}{G(Y_i)}f\{\theta(\bar{{X}}_i,\bbeta)|\bar X\}
\frac{\partial \theta(\bar{{X}}_i,\bbeta)}{\partial \bbeta}.
$$
(See the beginning of Section \ref{sec:asymptotics} for the definitions.)
Hence, $\partial \tilde{U}_n(\bbeta)/\partial \bbeta$ is continuous in
$\mathcal{B}$. Also, by the strong law of large numbers, the matrix $\partial
\tilde{U}_n(\bbeta)/\partial \bbeta$ converges almost surely
to
$A(\bbeta)=E\{ \partial /\partial \bbeta \tilde{U}_n(\bbeta)\}$
uniformly for $\bbeta \in \mathcal{B}$ as $n \rightarrow \infty$.
Finally,  it follows from the inverse function theorem that the unique solution
$\hat{\bbeta}$ converges to $\bbeta^o$ almost surely.

\mbox{}\par

\noindent\emph{Part 2:} Let
$$
U_n^G(\bbeta)=\frac{1}{n}\sum_{i=1}^n\frac{\Delta_i {Z}_i}{G(Y_i)}
\left( I\left[\int_0^{Y_i} \exp\{\bbeta' {X}_i(t)\} dt  > 1 \right]
-q \right).
$$
We write
\begin{eqnarray}
U_n(\bbeta^o)&=&U_n^G(\bbeta^o)+\left\{ U_n(\bbeta^o) - U_n^G(\bbeta^o)
\right\} \nonumber \\
&=& U_n^G(\bbeta^o)-\frac{1}{n}\sum_{i=1}^n\Delta_i
{Z}_i\frac{\hat{G}(Y_i)-G(Y_i)}{\hat{G}(Y_i)G(Y_i)}
\left(I\left[\int_0^{Y_i} \exp\{\bbeta^{o'} {X}_i(t)\} dt  > 1
\right] -q \right) \nonumber \\
&=& U_n^G(\bbeta^o) - \int_0^\infty \frac{\hat{G}(s)-G(s)}{\hat{G}(s)G(s)}
d {\Omega}(s) \nonumber
\end{eqnarray}
where
$$
{\Omega}(s)=\frac{1}{n}\sum_{i=1}^n {Z}_i
\left(I\left[\int_0^{s} \exp\{\bbeta^{o'} {X}_i(t)\} dt  > 1 \right]
-q \right) N_i(s)
$$
and $N_i(t)=I(Y_i \leq t) \Delta_i$.
Based on \citet[Theorem~3.2.3]{FH} we can show that
$\{\hat{G}(s)-G(s)\}/G(s)$ is asymptotically equivalent to $-\int_0^\infty
 dM_G(t)/R(t)$ where $M_G(t)=\sum_{i=1}^nM_{Gi}(t)$. Hence, by
interchanging the order of the integrals, we get that $U_n(\bbeta^o)$ is
asymptotically equivalent to
$$
U_n^G(\bbeta^o) + \int_0^\infty \left\{ \int_v^\infty
\frac{d{\Omega}(s)}{\hat{G}(s)} \right\} \frac{dM_G(v)}{R(v)}  .
$$
Since
$$
\int_v^\infty \frac{d{\Omega}(s)}{\hat{G}(s)} =
\frac{1}{n}\sum_{i=1}^n\frac{\Delta_i {Z}_i}{G(Y_i)}
\left(I\left[\int_0^{Y_i} \exp\{\bbeta^{o'} {X}_i(t)\} dt  > 1 \right]
-q \right)R_i(v)
+o_p(n^{-1/2+\delta}),
$$
we get
$$
\frac{1}{n}\int_0^\infty\left\{
\frac{1}{n^{-1}R(v)}\int_v^{\infty}\frac{d{\Omega}(s)}{G(s)}-\frac{u(\bbeta^o,v)}{r(v)}\right\}dM_G(v)
\rightarrow 0
$$
in probability as $n \rightarrow \infty$. Therefore, we conclude that $n^{1/2}
U_n(\bbeta^o)$
is asymptotically equivalent to $n^{-1/2} \sum_{i=1}^n \boleta_i$
where for $i=1,\ldots,n$:
$$
\boleta_{i}=\frac{\Delta_i {Z}_i}{G(Y_i)}
\left(I\left[\int_0^{Y_i} \exp\{\bbeta^{o'} {X}_i(t)\} dt  > 1
\right] -q\right)-\int_0^\infty \frac{u(\bbeta^o,v)}{r(v)}dM_{Gi}(v).
$$
Finally, by the multivariate central limit theorem we conclude that
$n^{1/2} U_n(\bbeta^o)$ converges weekly to a mean zero multivariate normal
distribution with covariance matrix $\Psi=E(\boleta_1\boleta_1')$.

\mbox{}

\noindent\emph{ Part 3:} Write
\begin{eqnarray}
&&\hspace{-1.5em}U_n(\bbeta)-U_n(\bbeta^o)
\\
&=&\frac{1}{n}\sum_{i=1}^n \frac{\Delta_i
{Z}_i}{G(Y_i)}\left( I\left[\int_0^{Y_i} \exp\{\bbeta'
{X}_i(t)\} dt  > 1 \right]  \nonumber  - I\left[\int_0^{Y_i} \exp\{\bbeta^{o'} {X}_i(t)\} dt  > 1
\right] \right) + o_p(n^{-1/2+\delta}) \nonumber \\
&=&\frac{1}{n}\sum_{i=1}^n \frac{\Delta_i {Z}_i}{G(Y_i)}\left(
I\left[\int_0^{Y_i} \exp\{\bbeta' {X}_i(t)\} dt  > 1 \right] - q
\right) + o_p(n^{-1/2}). \nonumber
\end{eqnarray}
Namely, $U_n(\bbeta)=U_n(\bbeta^o)+\tilde{U}_n(\bbeta)+o_p(n^{-1/2})$. By
Taylor expansion of $U_n(\hat{\bbeta})$ about $\bbeta^o$ we get
$$
0 \approx U_n(\bbeta^o) + \partial \tilde{U}_n(\bbeta) / \partial \bbeta
\mid_{\bbeta=\bbeta^o}(\hat{\bbeta}-\bbeta^o).
$$
Hence $n^{1/2}(\hat{\bbeta}-\bbeta^o)$ converges weakly to a zero mean normally
distributed random variable with variance $A(\bbeta^o)^{-1} \Psi
A(\bbeta^o)^{-1}$.

\subsection*{Proof of Theorem~\ref{thm2}}
\noindent\emph{ Part 1:}
Assume first that the censoring is independent of both covariates and failure time. Thus, $\hat G$ weakly converges to $ G$. Consequently, using the notation of $m(H_i,\bbeta)$ defined in~\eqref{eq:m}, we may write
\begin{align*}
 U_n^{DR}(\bbeta)= \frac{1}{n}\sum_{i=1}^n \left[ \frac{\Delta_i m(H_i, \bbeta)}{G_i(Y_i)}+\int_0^{\infty} Q\{t,\bbeta,\psi^*,H_i(t)\} \frac{dM_{Gi}(t)}{G(t)}\right]+o_p(n^{-1/2})\,
\end{align*}
by replacing the Kaplan-Meier estimator $\hat G$ with its limit, and $\hpsi$ with $\psi^*$.
Following Eq.~(3.10d) of \citet{robins_recovery_1992}, we can write $U_n^{DR}$ as
\begin{align}\label{eq:UNdr}
  U_n^{DR}(\bbeta)=\frac{1}{n}\sum_{i=1}^n \left(m(H_i,\bbeta)+\int_0^{\infty}\left[Q\{t,\bbeta,\psi^*,H_i(t)\} -m(H_i,\bbeta)\right]\frac{dM_{Gi}(t)}{G(t)}\right)+o_p(n^{-1/2})\,.
\end{align}
Since $M_{Gi}$, $i=1,\ldots,n$ are zero-mean martingales, $U_n^{DR}(\bbeta)$ converges to $E\{m(H,\bbeta)\}$ which has a unique zero at $\bo$ and hence $\tbeta$ is consistent.

Now assume that the posited model for $\{Z,T,\bX(T)\}$ holds. Then
\begin{align}\label{eq:posited_model_holds}
Q\{t,\bo,\psi^*,H(t)\}=E\left\{m(H,\bo)\left|T \geq t,H(t)\right.\right\}\,.
\end{align}
By~\eqref{eq:posited_model_holds}, for all $i$,
\begin{align*}
   R_i(t)Q\{t,\bo,\psi^*,H_i(t)\}&= R_i(t)E\left\{m(H_i,\bo)\left|T_i \geq t,H_i(t)\right.\right\}\\
  &= E\left\{R_i(t)m(H_i,\bo)\left|T_i \geq t,H_i(t)\right.\right\}\,,
\end{align*}
where the last equality follows from the Assumption~\ref{as:A6}. Similarly,
\begin{align*}
   (1-\Delta_i)Q\{C_i,\bo,\psi^*,H_i(C_i)\}&= (1-\Delta_i)E\left\{m(H_i,\bo)\left|T_i \geq C_i,H_i(C_i)\right.\right\}\\
  &= E\left\{(1-\Delta_i)m(H_i,\bo)\left|T_i> C_i,H_i(C_i)\right.\right\}\,.
\end{align*}

Write
$dM_{Gi}(t)=dN_{Gi}(t)-R_i(t)\lambda_G(t)dt$. Then, we may rewrite~\eqref{eq:UNdr} as
\begin{align*}
  U_n^{DR}(\bbeta)&= \frac{1}{n}\sum_{i=1}^n m(H_i,\bbeta)+\frac1n\sumin\frac{(1-\Delta_i)}{G(C_i)}\left[Q\{C_i,\bbeta,\psi^*,H_i(C_i)\}-m(H_i,\bbeta)\right]\\
  &-\frac1n\sumin\int_0^{\infty}\left[R_i(t)Q\{t,\bbeta,\psi^*,H_i(t)\} -R_i(t)m(H_i,\bbeta)\frac{\lambda_G(t)dt}{G(t)}\right]+o_p(n^{-1/2})\\
  &= \frac{1}{n}\sum_{i=1}^n m(H_i,\bbeta)\\
  &-\frac1n\sumin\frac{1}{G(C_i)}\left[(1-\Delta_i)m(H_i,\bbeta)-E\{(1-\Delta_i)m(H_i,\bbeta)|T_i>C_i,H_i(C_i)\}\right] \\
  &+\frac1n\sumin\int_0^{\infty}\left[R_i(t)m(H_i,\bbeta)-E\{R_i(t)m(H_i,\bbeta)|T_i\geq t, H_i(t)\}\right]\frac{\lambda_G(t)dt}{G(t)}+o_p(n^{-1/2})\\
  &\equiv \Upsilon_{1n}(\bbeta)+\Upsilon_{2n}(\bbeta)+\Upsilon_{3n}(\bbeta)+o_p(n^{-1/2})\,.
\end{align*}
By construction, $\Upsilon_{1n}(\bo)$ is $o_p(1)$. In the proof of Theorem~\ref{thm1}, we showed that $\Upsilon_{2n}(\bbeta)$ is in a Glivenko-Cantelli class. Using Corollary~9.27~(iii) of~\citet{Kosorok08}, we obtain that $\Upsilon_{3n}(\bbeta)$ is also in a Glivenko-Cantelli class. Since both $\Upsilon_{2n}(\bbeta)$ and $\Upsilon_{3n}(\bbeta)$ are the empirical means of mean zero random variables in a Glivenko-Cantelli class we have that
$\Upsilon_{2n}(\bbeta)+\Upsilon_{3n}(\bbeta)$ is $o_p(1)$ uniformly in $\bbeta$. Thus, $U_n^{DR}(\bbeta)$ converges to $E\{m(H,\bbeta)\}$ which has a unique zero at $\bo$ and hence $\tbeta$ is consistent.

\mbox{}

\noindent\emph{Part 2:} We now assume that the censoring is independent of both failure time and covariates. We already showed that
\begin{align*}
  n^{1/2}U_n(\bbeta^o)=n^{-1/2} \sum_{i=1}^n\left\{\frac{\Delta_i \bZ_i}{G(Y_i)}m(H_i,\bo)-\int_0^\infty \frac{u(\bbeta^o,v)}{r(v)}dM_{Gi}(v)\right\}+o_p(1)\,.
\end{align*}
Note that
\begin{align*}
  n^{-1/2}&\sum_{i=1}^n \int_0^{\infty}Q\{t,\bo,\hpsi,H_i(t)\} \frac{d{\hat M}_{{G}i}(t)}{\hat{G}(t)}
\\
   &=   n^{-1/2}\sum_{i=1}^n \int_0^{\infty}Q\{t,\bo,\psi^*,H_i(t)\} \frac{dM_{Gi}(t)}{G(t)}
\\
   &\quad+
  n^{-1/2}\sum_{i=1}^n \left[\int_0^{\infty}Q\{t,\bo,\hpsi,H_i(t)\}\frac{d{\hat M}_{Gi}(t)}{\hat G(t)}-
  \int_0^{\infty}Q\{t,\bo,\hpsi,H_i(t)\}\frac{dM_{Gi}(t)}{G(t)} \right]
\\
   &\quad+
  n^{-1/2}\sum_{i=1}^n \left[\int_0^{\infty}Q(t,\bo,\hpsi,H_i(t))\frac{dM_{Gi}(t)}{G(t)}-
  \int_0^{\infty}Q\{t,\bo,\psi^*,H_i(t)\}\frac{dM_{Gi}(t)}{G(t)} \right]
\\
      &=   n^{-1/2}\sum_{i=1}^n \int_0^{\infty}Q\{t,\bo,\psi^*,H_i(t)\} \frac{dM_{Gi}(t)}{G(t)}
\\
     &\quad+
  n^{1/2}(\hpsi-\psi^*)\frac{\partial}{\partial\psi}E\left[\int_0^{\infty}Q\{t,\bo,\psi^*,H(t)\}\frac{dM_{G1}(t)}{G(t)}\right]+o_p(1)\\
&= \quad  n^{-1/2}\sum_{i=1}^n \int_0^{\infty}Q\{t,\bo,\psi^*,H_i(t)\} \frac{dM_{Gi}(t)}{G(t)}+o_p(1)\,,
\end{align*}
where the one-before-last inequality follows from a first order Taylor expansion about $\psi^*$ and the mean zero martingale property. Noting that
\begin{align*}
u(\bbeta^o,v)&=\lim_{n \rightarrow \infty} \frac1n\sum_{i=1}^n\frac{\Delta_i
\bZ_i}{G(Y_i)}
\left(I\left[\int_0^{Y_i} \exp\{\bbeta^{o'} {X}_i(t)\} dt  > 1
\right] -q \right)I(Y_i \geq v),
\\
&=\lim_{n \rightarrow \infty} \frac{1}{n}\sum_{i=1}^n\frac{\Delta_i
\bZ_i}{G(T_i)}
\left(I\left[\int_0^{T_i} \exp\{\bbeta^{o'} {X}_i(t)\} dt  > 1
\right] -q \right)I(T_i \geq v)\\
&=E\{m(H,\bo)I(T\geq v)\},
\end{align*}
and $r(v)=\lim_{n \rightarrow \infty}n^{-1}R(v)=S(v)G(v)$, we obtain that
\begin{align*}
  \int_0^\infty \frac{u(\bbeta^o,v)}{r(v)}dM_{Gi}(v)=\int_0^\infty \frac{E\{m(H,\bo)I(T\geq v)\}}{S(v)}\frac{dM_{Gi}(v)}{G(v)}.
\end{align*}

Summarizing, we obtained that $n^{1/2}
U_n^{DR}(\bbeta^o)$
is asymptotically equivalent to $n^{-1/2} \sum_{i=1}^n \bxi_i$,
where for $i=1,\ldots,n$, $\bxi_i=\bxi_{i1}+\bxi_{i2}$,
$\bxi_{i1}=m(H_i,\bbeta)$, and
\begin{align*}
\bxi_{i2}&=\int_0^\infty\left[Q\{t,\bo,\psi^*,H_i(v)\}  -\frac{E\{m(H,\bo)I(T\geq v)\}}{S(v)}\right]dM_{Gi}(v).\\
\end{align*}

Finally, by the multivariate central limit theorem we conclude that
$n^{1/2} U_n^{DR}(\bbeta^o)$ converges weekly to a mean zero multivariate normal vector with covariance matrix $\Psi^{DR}=E(\bxi_1 \bxi_1')$.

\mbox{}

\noindent\emph{Part 3:} Write
\begin{align*}
U_n^{DR}(\bbeta)-U_n^{DR}(\bbeta^o)&=\frac1n\sum_{i=1}^n \frac{\Delta_i
\bZ_i}{G(Y_i)}\left( I\left[ \int_0^{Y_i} \exp\{\bbeta'
{X}_i(t)\} dt  > 1 \right] \right.  \\
& -\left. I\left[\int_0^{Y_i} \exp\{\bbeta^{o'} {X}_i(t)\} dt  > 1
\right] \right) + o_p(n^{-1/2}).
\end{align*}
Namely, $U_n^{DR}(\bbeta)=U_n^{DR}(\bbeta^o)+\tilde{U}_n(\bbeta)+o_p(n^{-1/2})$ where $\tilde{U}_n$ is defined in~\eqref{eq:Un_tilde}.
By
Taylor expansion of $U_n(\tbeta)$ about $\bbeta^o$ we get
$$
o_p(n^{-1/2})=U_n^{DR}(\tbeta)= U_n^{DR}(\bbeta^o) + \frac{\partial}{\partial \bbeta} \tilde{U}_n(\bbeta)
\mid_{\bbeta=\bbeta^o}(\tbeta-\bbeta^o)+o_p(n^{-1/2})\,.
$$
Hence $n^{1/2}(\tbeta-\bbeta^o)$ converges to a mean zero multivariate normal vector with variance $A(\bbeta^o)^{-1} \Psi^{DR}
A(\bbeta^o)^{-1}$.

\bibliographystyle{biometrika}
\bibliography{quantile}
\end{document}